\documentclass{ws-ijcia}
\usepackage{latexsym,graphics,pstricks}
\def\mathrm#1{{\mbox{\rm #1}}}

\title{OPEN-ENDED ARTIFICIAL EVOLUTION}
\author{RUSSELL K. STANDISH}

\address{High Performance Computing Support Unit, University of New South
  Wales\\2052, Australia\\
r.standish@unsw.edu.au, http://parallel.hpc.unsw.edu.au/rks}

\def\htmladdnormallink#1#2{#1}

\begin{document}
\maketitle
\begin{abstract}
  Of all the issues discussed at {\em Alife VII: Looking Forward,
    Looking Backward}, the issue of whether it was possible to create
  an artificial life system that exhibits {\em open-ended evolution} of
  novelty is by far the biggest. Of the 14 open problems settled on as
  a result of debate at the conference, some 6 are directly, or
  indirectly related to this issue.
  
  Most people equate open-ended evolution with complexity growth,
  although a priori these seem to be different things. In this paper I
  report on experiments to measure the complexity of Tierran
  organisms, and show the results for a {\em size-neutral} run of
  Tierra. In this run, no increase in organismal complexity was
  observed, although organism size did increase through the run. This
  result is discussed, offering some signposts on path to solving the
  issue of open ended evolution.
\end{abstract}

\section{Introduction}

An open discussion forum was conducted recently at Alife VII to draw
up a list of open problems that might guide the field of artificial
life in the coming years, styled on the famous Hilbert problems that
guided much of Mathematics in the 20th century. The resulting
list\cite{Bedau-etal2000} contains 14 open problems, divided into 3
categories: How does life arise from the non-living; What are the
potentials and limits of living systems; How is life related to mind,
machine and culture. 

The issue of {\em open-ended evolution} can be summed up by asking
under what conditions will an evolutionary system continue to produce
novel forms. Artificial Life systems such as Tierra and Avida produced
a rich diversity of organisms initially, yet ultimately peter
out.\cite{Adami98a} By contrast, the Earth's biosphere appears to
have continuously generated new and varied forms throughout the
$4\times10^9$ years of the history of life. There is also a clear
trend from simple organisms at the time of the first replicators
towards immensely complicated organisms such as mammals and birds
found on the Earth today. This raises the obvious question of {\em
  what is missing} in  artificial life systems? 

This issue is touched on most directly in problem 6 --- ``What is
inevitable in open-ended evolution of life'', but in fact is intimately
connected with problems 1, 2, 3, 5 and 7 also.

The issue of complexity growth is related to that of
open-endedness. Complexity is related to information in a direct
manner,\cite{Standish01a,Adami98a}  in a way to be made more precise
later in this paper. Loosely speaking, available complexity is
proportional to the dimension of phenotype space, and an evolutionary
process that remained at low levels of complexity will quickly exhaust
the possibilities for novel forms. However, intuitively, one would
expect the number of novel forms to increase exponentially with
available complexity, and so perhaps increasing complexity might cease
to be important factor in open-ended evolution beyond a certain
point. Of course, it is by far from proven that the number of possible
forms increases as rapidly with complexity as that, so it may still be
that complexity growth is essential for continual novelty. 

For the purposes of this paper, we will examine the possibilities for
open-ended growth in complexity in artificial life, since this most
closely resembles the evolution of our biosphere. It is worth bearing
in mind that the biosphere is subject to the {\em anthropic
principle}\cite{Standish00b} while no artificial life model is. Whether this
fact is important in the generation of complexity is a question worthy
of artificial life studies.

Results will also be reported of an experiment to measure the
complexity of Tierran organisms. Whilst it is widely believed that
the version of Tierra used here does not exhibit open ended complexity
growth, the Tierran team are hoping to produce a version of {\em
network Tierra} that does. The importance of this work is to
demonstrate the feasibility of these sorts of experiments on simple
ALife systems, before attempting more complex ones. 

\section{Complexity}

The concept of {\em complexity} as a measurable quantity has been
subject to a lot of debate. This is particularly due to the fact that
there are numerous ways of ordering things based on physical
quantities that agree with our subjective notion of complexity, within
particular application environments. Examples include organism size,
numbers of parts and number of different cell types in the arena of
complexity of animals, to name just a few.\cite{McShea96} A good
discussion of these different types can be found in
Edmonds.\cite{Edmonds99} Unfortunately, these measures typically do
not extend to other application areas, where most of the complexity
may be encoded in the connections between the parts, or in the levels of
hierarchy.

There is, however, a unifying concept that covers all notions of
measurable complexity, and that is {\em information}.  Information was
first quantified in the landmark work of Shannon\cite{Shannon49} in the
late 1940s, however in this case the semantic content of information
was ignored, since the purpose was to measure the capacities of
communication channels. The next theoretical step occurred in the
1960s with the advent of {\em algorithmic complexity}, also known as
{\em Kolmogorov complexity}.\cite{Kolmogorov65,Chaitin66,Solomonoff64} In
algorithmic complexity, the complexity of a description is defined to be
the length of the shortest program that can generate the description,
when run on a particular {\em universal Turing machine} (UTM). This
complexity is closely related to the Shannon entropy in the following
way. Consider the set of all infinite length descriptions expressed as binary
strings. A halting program of length $\ell$ will be equivalent to all
infinite length descriptions having the same prefix of length $\ell$,
and so will have probability $2^{-\ell}$ of being picked at random
from the set of all descriptions. This gives rise to the so called
{\em universal prior measure}, defined as\cite{Li-Vitanyi97}
\begin{equation}\label{Universal Prior}
P_U(x)=\sum_{p:U \mbox{\rm\scriptsize\ computes } x \mbox{\rm\scriptsize\ from } p \mbox{\rm\scriptsize\ and
    halts}} 2^{-\ell(p)},
\end{equation}
where $U$ is the reference universal Turing machine, and $\ell(p)$ is
the length of program $p$. The Shannon entropy computed using this
probability distribution differs from the algorithmic complexity by at
most a constant that depends only on the Turing machines chosen, not
the description.\cite[Coding Theorem 4.3.3]{Li-Vitanyi97}
\begin{equation}\label{UTM complexity}
C(x) \equiv -\log_2 P_U(x) = K_U(x) + C_U, \,\, \exists C_U\geq0
\end{equation}
where $K_U(x)$ is the algorithmic complexity of $x$.

There are two key problems with algorithmic complexity. The first is
the arbitrariness of the reference Turing machine $U$. The second
relates to the behaviour of random strings --- the algorithmic
complexity of a random string is at least as great as the string
length itself, ie has maximal complexity. This is in conflict with our
common sense notion of random strings having no information at
all.\cite{Gell-Mann94}

The first problem is really a non-problem. All information is {\em
  context dependent}, with the context being defined by the receiver
or interpreter of a message. See Adami\cite{Adami98a} for an
explicit discussion of this issue. In the case of algorithmic
complexity, the interpreter is the reference UTM, which must be agreed
upon in the context of the discussion.

The second issue can be dealt with by broadening the concept of
algorithmic complexity in the following way:\cite{Standish01a} given
an interpreter of information that defines the context, the
interpreter must classify all possible messages into a countable,
possibly finite set of valid meanings. The UTM classifies messages
into equivalent programs. A human interpreter will have a somewhat
different scheme of classifying things. With respect to the particular
issue of random strings, a human being will classify all random
strings as identical --- all gibberish. The obvious generalisation of
complexity is:
\begin{equation}\label{complexity}
C(x) = \lim_{s\rightarrow\infty} s\log_2 N - \log_2 \omega(s,x)
\end{equation}
where $C(x)$ is the complexity (measured in bits), $N$ the size of the
alphabet used to encode the description and $\omega(s,x)$ the size of
the class of all descriptions equivalent to $x$ and of length less
than $s$. This definition allows for infinite strings to be measured,
provided the above limit exists. Random strings are dense in the set
of all possible strings, so $\omega(s,x) \rightarrow N^s$ as
$s\rightarrow\infty$. Random strings have zero complexity.

To see that this definition of $C(x)$ converges for finite
descriptions (ie ones for which the interpreter ``halts'' and gives an
interpretation after a fixed number $n$ characters), note that all
strings having the same $n$ character prefix are equivalent.
Therefore, $\omega(s,x) \geq N^{s-n}$, so $C(x) \leq n\log_2N$.
Equality only holds iff $x$ is incompressible.

In the case where the interpreter is the UTM $U$, each finite
prefix bitstring $p$ equivalent to $x$ contributes $2^{s-\ell(p)}$ to
$\omega(s,x)$. So 
\begin{eqnarray*}
\omega(s,x) &=& \sum_{p:U \mbox{\rm\scriptsize\ computes } x \mbox{\rm\scriptsize\ from } p \mbox{\rm\scriptsize\ and
    halts}} 2^{s-\ell(p)} = 2^sP_U(x)\;\; \\
&\Rightarrow& \hspace{5em} C(x) = -\lim_{s\rightarrow\infty} P_U(x)
\end{eqnarray*}
so $C(x)$ reduces to the usual complexity measure in equation
(\ref{UTM complexity}).

\section{Complexity in Tierra}

Now that complexity has been defined in a general sense, we need to
work out an appropriate context for discussing the complexity of
Tierran organisms. Tierra is an artificial life system created by Tom
Ray\cite{Ray91} where the artificial organisms are computer programs
written in a special assembler-like language. The equivalence
principle for defining the context is phenotypic equivalence, ie when
two Tierran organisms behave identically, even if they contain a
different sequence of instructions. Adami's group pioneered this
technique with {\em Avida},\cite{Adami98a,Adami-Cerf00} a similar
artificial life system inspired by Tierra. In Avida, the situation is
particularly simple, as organisms have only one phenotypic
characteristic, namely their reproduction rate. To make matters more
interesting, the Avida group set the creatures a range of
computational tasks to attempt --- success at these tasks is rewarded
by extra CPU resources. The success or otherwise of the creatures at
these tasks further distinguishes between Avida phenotypes.

Tierra presents considerably greater difficulties than Avida in that
Tierran organisms can {\em interact} with each other via template
matches or simply ``falling off the end of the code''. So, the Tierra
phenotype can be categorised by examining the interactions between all
possible phenotypes. Since genotype space is so huge, what better way
of searching for viable phenotypes than using a genetic algorithm,
namely taking the results from a long Tierra run. These are then
pitted against each other in pairwise
tournaments\cite{Standish99a,Standish97b} in a specially crafted
simulator called {\em miniTierra}\cite{Standish97a} that can perform
the tournament nearly 1000 times faster than Tierra itself. Once all
the tournaments have been completed, the organisms are sorted into
phenotypically equivalent classes, resulting in a small list of 26
archetypal organisms. These are labeled by the genotype with the
earliest creation date. Table \ref{archetypes} lists the archetypes,
along with their creation time, non-volatile site count (to be explained
later) and measured entropy.

Having now established an archetype list, we need to estimate the
density of neutrally equivalent genotypes in genotype space in order
to apply the complexity formula (\ref{complexity}). Unfortunately,
simply sampling the whole of genotype space is computation infeasible,
as the overall density is low --- an organism having complexity of 60
instructions (300 bits) will have a density of $32^{-60}$ requiring
many more samples than there are nanoseconds in the history of the
universe\footnote{Tierra uses 5 bit instructions, so there are 32
  possible instructions at each site of the digital genome}. We can,
however, explore the {\em neutral network} of phenotypic equivalents
connected by single mutations. Unfortunately, this also is
computationally infeasible --- take our previous example of an
organism of length 80 instructions, having a complexity of 60
instructions. The neutrally equivalent set will have size $32^{20}$
genotypes, each of which needs to be checked.

What we can do is a hybrid method: begin by exploring the neutral
network out to a certain cutoff value (usually 2 hops in this
experiment, however several of the cheaper organisms were computed to
higher cutoff value to check on convergence). Where the network is
thickly branching, perform a Monte-Carlo style sampling of that
subspace, since this is likely to correspond to part of the genome
that contains little information. Where the network has only a few
branches, then continue following the network branches. The threshold
between the two strategies is determined by some heuristic estimates
of how expensive each approach is likely to be. Basically the density
of neutral variants in a genotype subspace is assumed to be
proportional to the number of immediate neutral neighbours in that
subspace. This then entails both the cost of traversing the network
(which rises exponentially with neutral density) and the cost of
performing Monte Carlo sampling, which falls with increasing neutral
density. 

The code and data used for this experiment is available from \linebreak
\htmladdnormallink{http://parallel.hpc.unsw.edu.au/rks/software/eco-tierra-1.1.tar.gz}{http://parallel.hpc.unsw.edu.au/rks/software/eco-tierra-1.1.tar.gz}.
The use of the software is briefly documented in the README file.

We end up with a probabilistic estimate of the density of
phenotypic equivalents, and hence the complexity by
(\ref{complexity}). This estimate should converge as more of the
neutral network is searched.

Since the first thing to compute is the number of nearest neutral
neighbours at each site, a simple complexity estimate can be found: classify
sites as hot if every possible mutation leaves the organism neutrally
equivalent, and cold if any mutation changes the phenotype. The number
of cold sites could be called the {\em non-volatile site count},
denoted $C_\mathrm{NV}$. A slightly more sophisticated calculation is
to compute
\begin{equation}\label{sse}
C_\mathrm{ss}=\ell-\sum_{i=1}\ell\log_{32}n_i
\end{equation}
where $n_i$ is the number of mutations at site $i$ on the genome that
lead to differing phenotypes, which we can call the {\em single site
entropy}. For a completely cold site, $n_i=1$, and for a hot site
$n_i=32$. Formula (\ref{sse}) takes into account shades of grey in
``site temperature''. Adami {\em et al.} use this measurement in their
work.\cite{Adami98a} Equation (\ref{sse}) is clearly an underestimate
of the amount of complexity defined in (\ref{complexity}), as is the
non-volatile site count.

\section{Results}

\begin{figure}
\setlength{\unitlength}{0.240900pt}
\ifx\plotpoint\undefined\newsavebox{\plotpoint}\fi
\sbox{\plotpoint}{\rule[-0.200pt]{0.400pt}{0.400pt}}%
\begin{picture}(1500,900)(0,0)
\font\gnuplot=cmr10 at 10pt
\gnuplot
\sbox{\plotpoint}{\rule[-0.200pt]{0.400pt}{0.400pt}}%
\put(161.0,123.0){\rule[-0.200pt]{4.818pt}{0.400pt}}
\put(141,123){\makebox(0,0)[r]{0}}
\put(1419.0,123.0){\rule[-0.200pt]{4.818pt}{0.400pt}}
\put(161.0,205.0){\rule[-0.200pt]{4.818pt}{0.400pt}}
\put(141,205){\makebox(0,0)[r]{50}}
\put(1419.0,205.0){\rule[-0.200pt]{4.818pt}{0.400pt}}
\put(161.0,287.0){\rule[-0.200pt]{4.818pt}{0.400pt}}
\put(141,287){\makebox(0,0)[r]{100}}
\put(1419.0,287.0){\rule[-0.200pt]{4.818pt}{0.400pt}}
\put(161.0,368.0){\rule[-0.200pt]{4.818pt}{0.400pt}}
\put(141,368){\makebox(0,0)[r]{150}}
\put(1419.0,368.0){\rule[-0.200pt]{4.818pt}{0.400pt}}
\put(161.0,450.0){\rule[-0.200pt]{4.818pt}{0.400pt}}
\put(141,450){\makebox(0,0)[r]{200}}
\put(1419.0,450.0){\rule[-0.200pt]{4.818pt}{0.400pt}}
\put(161.0,532.0){\rule[-0.200pt]{4.818pt}{0.400pt}}
\put(141,532){\makebox(0,0)[r]{250}}
\put(1419.0,532.0){\rule[-0.200pt]{4.818pt}{0.400pt}}
\put(161.0,614.0){\rule[-0.200pt]{4.818pt}{0.400pt}}
\put(141,614){\makebox(0,0)[r]{300}}
\put(1419.0,614.0){\rule[-0.200pt]{4.818pt}{0.400pt}}
\put(161.0,695.0){\rule[-0.200pt]{4.818pt}{0.400pt}}
\put(141,695){\makebox(0,0)[r]{350}}
\put(1419.0,695.0){\rule[-0.200pt]{4.818pt}{0.400pt}}
\put(161.0,777.0){\rule[-0.200pt]{4.818pt}{0.400pt}}
\put(141,777){\makebox(0,0)[r]{400}}
\put(1419.0,777.0){\rule[-0.200pt]{4.818pt}{0.400pt}}
\put(161.0,123.0){\rule[-0.200pt]{0.400pt}{4.818pt}}
\put(161,82){\makebox(0,0){0}}
\put(161.0,757.0){\rule[-0.200pt]{0.400pt}{4.818pt}}
\put(374.0,123.0){\rule[-0.200pt]{0.400pt}{4.818pt}}
\put(374,82){\makebox(0,0){1000}}
\put(374.0,757.0){\rule[-0.200pt]{0.400pt}{4.818pt}}
\put(587.0,123.0){\rule[-0.200pt]{0.400pt}{4.818pt}}
\put(587,82){\makebox(0,0){2000}}
\put(587.0,757.0){\rule[-0.200pt]{0.400pt}{4.818pt}}
\put(800.0,123.0){\rule[-0.200pt]{0.400pt}{4.818pt}}
\put(800,82){\makebox(0,0){3000}}
\put(800.0,757.0){\rule[-0.200pt]{0.400pt}{4.818pt}}
\put(1013.0,123.0){\rule[-0.200pt]{0.400pt}{4.818pt}}
\put(1013,82){\makebox(0,0){4000}}
\put(1013.0,757.0){\rule[-0.200pt]{0.400pt}{4.818pt}}
\put(1226.0,123.0){\rule[-0.200pt]{0.400pt}{4.818pt}}
\put(1226,82){\makebox(0,0){5000}}
\put(1226.0,757.0){\rule[-0.200pt]{0.400pt}{4.818pt}}
\put(1439.0,123.0){\rule[-0.200pt]{0.400pt}{4.818pt}}
\put(1439,82){\makebox(0,0){6000}}
\put(1439.0,757.0){\rule[-0.200pt]{0.400pt}{4.818pt}}
\put(161.0,123.0){\rule[-0.200pt]{307.870pt}{0.400pt}}
\put(1439.0,123.0){\rule[-0.200pt]{0.400pt}{157.549pt}}
\put(161.0,777.0){\rule[-0.200pt]{307.870pt}{0.400pt}}
\put(40,450){\makebox(0,0){\rotateleft{Instructions}}}
\put(800,21){\makebox(0,0){Millions of timesteps}}
\put(800,839){\makebox(0,0){Computed complexities of Tierran Organisms}}
\put(161.0,123.0){\rule[-0.200pt]{0.400pt}{157.549pt}}
\put(1279,737){\makebox(0,0)[r]{$C_\mathrm{NV}$}}
\put(1248,157){\raisebox{-.8pt}{\makebox(0,0){$\Diamond$}}}
\put(161,179){\raisebox{-.8pt}{\makebox(0,0){$\Diamond$}}}
\put(837,170){\raisebox{-.8pt}{\makebox(0,0){$\Diamond$}}}
\put(161,226){\raisebox{-.8pt}{\makebox(0,0){$\Diamond$}}}
\put(174,210){\raisebox{-.8pt}{\makebox(0,0){$\Diamond$}}}
\put(918,183){\raisebox{-.8pt}{\makebox(0,0){$\Diamond$}}}
\put(161,224){\raisebox{-.8pt}{\makebox(0,0){$\Diamond$}}}
\put(162,216){\raisebox{-.8pt}{\makebox(0,0){$\Diamond$}}}
\put(798,167){\raisebox{-.8pt}{\makebox(0,0){$\Diamond$}}}
\put(187,157){\raisebox{-.8pt}{\makebox(0,0){$\Diamond$}}}
\put(223,187){\raisebox{-.8pt}{\makebox(0,0){$\Diamond$}}}
\put(224,190){\raisebox{-.8pt}{\makebox(0,0){$\Diamond$}}}
\put(201,152){\raisebox{-.8pt}{\makebox(0,0){$\Diamond$}}}
\put(199,152){\raisebox{-.8pt}{\makebox(0,0){$\Diamond$}}}
\put(216,200){\raisebox{-.8pt}{\makebox(0,0){$\Diamond$}}}
\put(173,210){\raisebox{-.8pt}{\makebox(0,0){$\Diamond$}}}
\put(175,130){\raisebox{-.8pt}{\makebox(0,0){$\Diamond$}}}
\put(175,210){\raisebox{-.8pt}{\makebox(0,0){$\Diamond$}}}
\put(247,185){\raisebox{-.8pt}{\makebox(0,0){$\Diamond$}}}
\put(701,169){\raisebox{-.8pt}{\makebox(0,0){$\Diamond$}}}
\put(685,167){\raisebox{-.8pt}{\makebox(0,0){$\Diamond$}}}
\put(679,170){\raisebox{-.8pt}{\makebox(0,0){$\Diamond$}}}
\put(230,201){\raisebox{-.8pt}{\makebox(0,0){$\Diamond$}}}
\put(237,201){\raisebox{-.8pt}{\makebox(0,0){$\Diamond$}}}
\put(866,170){\raisebox{-.8pt}{\makebox(0,0){$\Diamond$}}}
\put(868,169){\raisebox{-.8pt}{\makebox(0,0){$\Diamond$}}}
\put(1349,737){\raisebox{-.8pt}{\makebox(0,0){$\Diamond$}}}
\put(1279,696){\makebox(0,0)[r]{$C_2$}}
\put(1248,166){\makebox(0,0){$+$}}
\put(161,185){\makebox(0,0){$+$}}
\put(837,182){\makebox(0,0){$+$}}
\put(161,236){\makebox(0,0){$+$}}
\put(174,224){\makebox(0,0){$+$}}
\put(918,195){\makebox(0,0){$+$}}
\put(161,237){\makebox(0,0){$+$}}
\put(162,231){\makebox(0,0){$+$}}
\put(798,180){\makebox(0,0){$+$}}
\put(187,169){\makebox(0,0){$+$}}
\put(223,205){\makebox(0,0){$+$}}
\put(224,208){\makebox(0,0){$+$}}
\put(201,166){\makebox(0,0){$+$}}
\put(199,166){\makebox(0,0){$+$}}
\put(216,215){\makebox(0,0){$+$}}
\put(173,224){\makebox(0,0){$+$}}
\put(175,134){\makebox(0,0){$+$}}
\put(175,224){\makebox(0,0){$+$}}
\put(247,205){\makebox(0,0){$+$}}
\put(701,187){\makebox(0,0){$+$}}
\put(685,185){\makebox(0,0){$+$}}
\put(679,187){\makebox(0,0){$+$}}
\put(230,216){\makebox(0,0){$+$}}
\put(237,216){\makebox(0,0){$+$}}
\put(866,182){\makebox(0,0){$+$}}
\put(868,180){\makebox(0,0){$+$}}
\put(1349,696){\makebox(0,0){$+$}}
\sbox{\plotpoint}{\rule[-0.400pt]{0.800pt}{0.800pt}}%
\put(1279,655){\makebox(0,0)[r]{length}}
\put(1248,185){\raisebox{-.8pt}{\makebox(0,0){$\Box$}}}
\put(161,197){\raisebox{-.8pt}{\makebox(0,0){$\Box$}}}
\put(837,229){\raisebox{-.8pt}{\makebox(0,0){$\Box$}}}
\put(161,242){\raisebox{-.8pt}{\makebox(0,0){$\Box$}}}
\put(174,251){\raisebox{-.8pt}{\makebox(0,0){$\Box$}}}
\put(918,251){\raisebox{-.8pt}{\makebox(0,0){$\Box$}}}
\put(161,254){\raisebox{-.8pt}{\makebox(0,0){$\Box$}}}
\put(162,254){\raisebox{-.8pt}{\makebox(0,0){$\Box$}}}
\put(798,255){\raisebox{-.8pt}{\makebox(0,0){$\Box$}}}
\put(187,332){\raisebox{-.8pt}{\makebox(0,0){$\Box$}}}
\put(223,339){\raisebox{-.8pt}{\makebox(0,0){$\Box$}}}
\put(224,342){\raisebox{-.8pt}{\makebox(0,0){$\Box$}}}
\put(201,349){\raisebox{-.8pt}{\makebox(0,0){$\Box$}}}
\put(199,349){\raisebox{-.8pt}{\makebox(0,0){$\Box$}}}
\put(216,350){\raisebox{-.8pt}{\makebox(0,0){$\Box$}}}
\put(173,368){\raisebox{-.8pt}{\makebox(0,0){$\Box$}}}
\put(175,376){\raisebox{-.8pt}{\makebox(0,0){$\Box$}}}
\put(175,380){\raisebox{-.8pt}{\makebox(0,0){$\Box$}}}
\put(247,421){\raisebox{-.8pt}{\makebox(0,0){$\Box$}}}
\put(701,427){\raisebox{-.8pt}{\makebox(0,0){$\Box$}}}
\put(685,440){\raisebox{-.8pt}{\makebox(0,0){$\Box$}}}
\put(679,447){\raisebox{-.8pt}{\makebox(0,0){$\Box$}}}
\put(230,479){\raisebox{-.8pt}{\makebox(0,0){$\Box$}}}
\put(237,509){\raisebox{-.8pt}{\makebox(0,0){$\Box$}}}
\put(866,548){\raisebox{-.8pt}{\makebox(0,0){$\Box$}}}
\put(868,772){\raisebox{-.8pt}{\makebox(0,0){$\Box$}}}
\put(1349,655){\raisebox{-.8pt}{\makebox(0,0){$\Box$}}}
\end{picture}
\caption{Plot of the results for the archetypes. The horizontal axis
  plots the time of first appearance, and for each organism is plotted
  the length, non-volatile site count $C_\mathrm{NV}$ and the
  complexity $C_2$.}
\label{tierra-info-fig}
\end{figure}
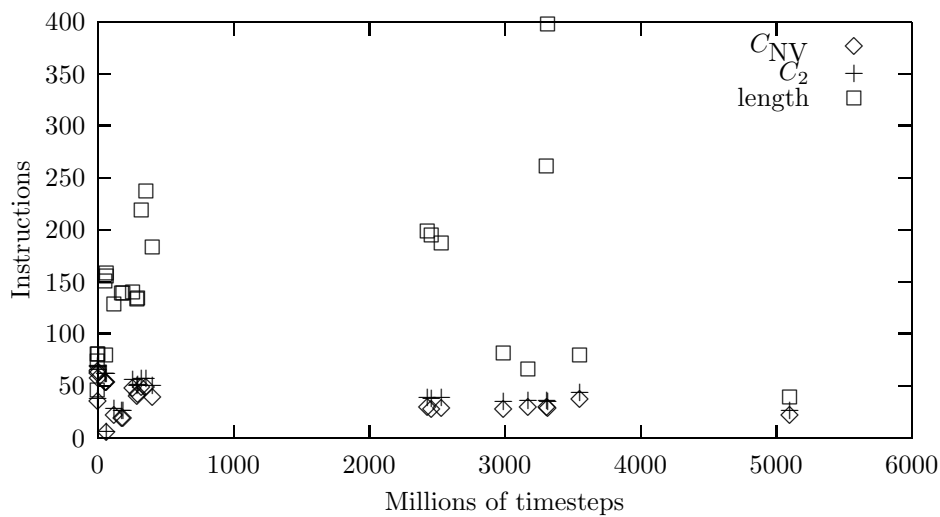

Table \ref{archetypes} shows the creation time (in millions of
instructions), computed single site entropies and complexity measures
estimated by truncating the neutral network at two hops (denoted
$C_2$) for each of the phenotypes found in this sample run. Figure
\ref{tierra-info-fig} shows the same results in graphical form, with
the creation time plotted along the horizontal axis, and the organism
length, non-volatile site count $C_\mathrm{NV}$ and $C_2$ plotted as
separate points. There are several key points to note:
\begin{itemize}
\item Over this Tierra run of $6\times10^9$ instructions, there is no
  sign of complexity increase, although lengths of the organisms do
  increase from 80 initially to 526 instructions long by
  $1.4\times10^9$ instructions executed. CPU resources are divided
  amongst the organisms in a {\em size neutral} fashion (ie
  proportional to organism size) to allow this growth in genome size
  to happen. This is not an especially lengthy Tierra run, however,
  and perhaps real complexity takes much longer accumulate in the junk
  parts of the genome.
\item The non-volatile site count $C_\mathrm{NV}$ is a good proxy
  measure of the total complexity. The biggest discrepancy in this
  database occurred for 0194aag, and the real complexity was only 58\%
  higher. This is good news, as it is far more tractable to compute
  $C_\mathrm{NV}$, than the full complexity. The full $C_2$
  computation for this dataset of organisms required nearly 8 CPU
  years of contemporary processor time, as opposed to about 3 CPU
  hours for $C_\mathrm{NV}$. Fortunately, it is a highly parallel
  problem, and was computed within about 3 months on a major high
  performance computing facility.
\item The precise definition of the {\em phenotype} has a big
  influence over complexity values. As has been argued by
  Adami\cite{Adami98a} and myself,\cite{Standish01a} this is highly
  context dependent. No particular definition is wrong {\em per se}, however
  some definitions will be better than others. I hope, dear reader,
  you consider my definition based on pairwise ecological interactions
  to be a good definition, and that therefore the complexity measures
  obtained are useful. It is known that Tierran organisms can exhibit
  what is called {\em social hyperparasitism}.\cite{Ray91} This is
  necessarily a 3-way interaction between Tierran organisms --- {\em
    hyperparasitism} refers to stealing another organism's CPU to
  reproduce itself, and {\em social} refers to the fact that multiple
  organisms must cooperate to perform this feat. However both of these
  properties show up as distinct patterns amongst 2 way interactions -
  the hyperparasite still manages to interfere with its prey's
  reproductive capability, even if it is unable to gain from the act
  alone, and mutual interaction with members of its own kind show up
  also.

\end{itemize}

\begin{table}
\begin{tabular}{|l|l|l|l||l|l|l|l|}
\hline
Organism & Creation & $C_\mathrm{NV}$ & $C_2$ & Organism & Creation & $C_\mathrm{NV}$ & $C_2$\\ 
\hline
0038aep & 5104 & 21   &   26.2362 & 0138aai & 179  & 18   &   26.1733  \\ 
0045aaa & 2    & 34   &   36.8368 & 0139aaa & 258  & 47   &   62.8791  \\
0065aac & 3172 & 29   &   35.5448 & 0150aag & 55   & 53   &   62.8791  \\
0073aaa & 0    & 63   &   65.2125 & 0155aab & 66   & 4    &   5        \\
0078aal & 59   & 53   &   63.0305 & 0157aaa & 66   & 53   &   62.9423  \\
0078aan & 3556 & 37   &   43.5753 & 0182aaa & 402  & 38   &   49.1947  \\
0080aaa & 0    & 62   &   69.9581 & 0186aah & 2536 & 28   &   41.2842  \\
0080aea & 3    & 57   &   66.1046 & 0194aag & 2461 & 27   &   42.7729  \\
0081aaj & 2989 & 27   &   32.7449 & 0198aad & 2434 & 29   &   39.3959  \\
0128aad & 123  & 21   &   27.8352 & 0218aab & 323  & 48   &   57.3709  \\
0132abi & 289  & 39   &   50.0246 & 0236aaa & 359  & 48   &   63.836   \\
0134aae & 298  & 41   &   51.6064 & 0260aae & 3310 & 29   &   35.8535  \\
0138aab & 190  & 18   &   25.9443 & 0397aab & 3321 & 28   &   34.2954  \\
\hline
\end{tabular}
\caption{Table of archetypal organisms, with the time of first
appearance in genebank record, their non-volatile site count
$C_\mathrm{NV}$ and their computed complexity truncating the neutral
network at 2 hops $C_2$.}
\label{archetypes}
\end{table}

\section{Discussion}

Thus far, the focus of this work has been on presenting a practical
method of computing the complexity of Tierran organisms. However, it
has been long recognised that neither Tierra nor Avida exhibit
open-ended evolution evolution. It is thought that these systems
rapidly adapt to and exhaust the possibilities of a fairly simple
environment. Adami {\em et al.} have approached the problem by artificially
increasing the complexity of the environment by specifying a sequence
of arithmetical operations that the digital organisms can
attempt. Success is rewarded by extra CPU time. In these
circumstances, they have shown increasing complexity of the organisms
over time.\cite{Adami98a,Adami-Cerf00} Ray, on the other hand, is
attempting to exploit naturally occurring environmental complexity
provided by the Internet,\cite{Ray-reserve} and hopes to engineer a
``Cambrian explosion''. To date, his team has had mixed success ---
whilst they have managed to get organisms to persist in a
multicellular state under mutational load, as opposed to reverting to
a single celled, they haven't achieved their aim of complexity
growth.\cite{Ray-Hart98}

The obvious next step in research is to attempt to measure the
complexities of network Tierra organisms. It is by no means clear that
this is tractable. The success of this approach with the original
Tierran and Avidan organisms may well be due to the simplicity of the
environment.

As I observed,\cite{Standish00b} the natural biosphere operates
under the constraints of the {\em anthropic principle}. This means that
we must necessarily observe a path to increased complexity in the
biosphere. Furthermore, there are some statistical regularities in
evolutionary history that indicates the evolution of the biosphere to
be an extremely rare process rather than an inevitable
one.\cite{Hanson00} The universe appears to be performing a quantum
search algorithm to find the precise conditions required to generate
intelligent life. Probably the way forward is to perform a similar
sort of ``scattergun'' approach, using the power of quantum computers
if and when they become available.

A key step in doing this is to generate a process that adaptively
recognises complexity, since it will be impossible to include humans
in the loop, even when run on conventional computing platforms. To
this end, techniques developed for {\em data mining}\cite{Marakas99}
should prove useful. As a curious twist, artificial life techniques
are being successfully applied to the domain of data
mining.\cite{Monmarche-etal99} Developing a coevolving system observer
is something that can be started now with present technology.

\section{Acknowledgments}

The author would like to thank the {\em Australian Centre for Advanced
Computing and Communications} (ac3) and the {\em Australian
Partnership for Advanced Computing} for generous grants of computing
resources, without which this project would not have been feasible.


\begin{thebibliography}{10}

\bibitem{Adami98a}
Chris Adami.
\newblock {\em Introduction to Artificial Life}.
\newblock Springer, 1998.

\bibitem{Adami-Cerf00}
Chris Adami and N.J.Cerf.
\newblock Physical complexity of symbolic sequences.
\newblock {\em Physica D}, 137:62--69, 2000.

\bibitem{Bedau-etal2000}
Mark~A. Bedau, John~S. McCaskill, Norman~H. Packard, Steen Rasmussen, Chris
  Adami, Devid~G. Green, Takashi Ikegami, Kinihiko Kaneko, and Thomas~S. Ray.
\newblock Open problems in artificial life.
\newblock {\em Artificial Life}, 6:363--376, 2000.

\bibitem{Chaitin66}
G.~J. Chaitin.
\newblock On the length of programs for computing finite binary sequences.
\newblock {\em J. Assoc. Comput. Mach.}, 13:547--569, 1966.

\bibitem{Edmonds99}
B.~Edmonds.
\newblock {\em Syntactic Measures of Complexity}.
\newblock PhD thesis, University of Manchester, 1999.
\newblock http://www.cpm.mmu.ac.uk/\~{}bruce/thesis.

\bibitem{Floreano-etal99}
Dario Floreano, Jean-Daniel Nicoud, and Francesco Mondada, editors.
\newblock {\em Advances in Artificial Life: 5th European Conference, ECAL 99},
  volume 1674 of {\em Lecture Notes in Computer Science}.
\newblock Springer, Berlin, 1999.

\bibitem{Gell-Mann94}
M.~Gell-Mann.
\newblock {\em The Quark and the Jaguar: Adventures in the Simple and the
  Complex}.
\newblock Freeman, 1994.

\bibitem{Hanson00}
Robin Hanson.
\newblock Must early life be easy? the rhythm of major evolutionary transitions.
\newblock {\em Origins of Life}, 2000.
\newblock submitted; http://hanson.gmu.edu/hardstep.pdf.

\bibitem{Kolmogorov65}
A.~N. Kolmogorov.
\newblock Three approaches to the quantitative definition of information.
\newblock {\em Problems Information Transmission}, 1:1--7, 1965.

\bibitem{Li-Vitanyi97}
Ming Li and Paul Vit\'anyi.
\newblock {\em An Introduction to {Kolmogorov} Complexity and its
  Applications}.
\newblock Springer, New York, 2nd edition, 1997.

\bibitem{Marakas99}
George~M. Marakas.
\newblock {\em Decision Support Systems in the Twenty-First Century}.
\newblock Prentice Hall, Upper Saddle River, NJ, 1999.

\bibitem{McShea96}
Daniel~W. McShea.
\newblock Metazoan complexity and evolution: Is there a trend?
\newblock {\em Evolution}, 50:477--492, 1996.

\bibitem{Monmarche-etal99}
N.~Monmarch\'e, M.~Slimane, and G.~Venturini.
\newblock On improving clustering in numerical databases with artificial ants.
\newblock In Floreano et~al. \cite{Floreano-etal99}, pages 626--635.

\bibitem{Ray-reserve}
Tom Ray.
\newblock A proposal to create two biodiversity reserves: One digital and one
  organic.
\newblock See ftp://tierra.slhs.udel.edu/tierra/doc/reserves.tex,
  http://www.hip.atr.co.jp/\~{}ray/pubs/reserves/reserves.html. Also see New
  Scientist, vol 150, no 2034, pp32--35.

\bibitem{Ray91}
Tom Ray.
\newblock An approach to the synthesis of life.
\newblock In C.~G. Langton, C.~Taylor, J.~D. Farmer, and S.~Rasmussen, editors,
  {\em Artificial Life II}, page 371. Addison-Wesley, New York, 1991.

\bibitem{Ray-Hart98}
Tom Ray and Joseph Hart.
\newblock Evolution of differentiated multi-threaded digital organisms.
\newblock In Chris Adami, Richard Belew, Hiroaki Kitano, and Charles Taylor,
  editors, {\em Artificial Life VI}, pages 295--304, Cambridge, Mass., 1998.
  MIT Press.

\bibitem{Shannon49}
Claude~E. Shannon.
\newblock {\em The Mathematical Theory of Communication}.
\newblock Urbana, 1949.

\bibitem{Solomonoff64}
R.~J. Solomonoff.
\newblock A formal theory of inductive inference, part 1 and 2.
\newblock {\em Inform. Contr.}, pages 1--22,224--254, 1964.

\bibitem{Standish97b}
R.~K. Standish.
\newblock Embryology in {T}ierra: A study of a genotype to phenotype map.
\newblock {\em Complexity International}, 4, 1997.

\bibitem{Standish97a}
R.~K. Standish.
\newblock On an efficient implementation of {T}ierra.
\newblock {\em Complexity International}, 4, 1997.

\bibitem{Standish99a}
Russell~K. Standish.
\newblock Some techniques for the measurement of complexity in {Tierra}.
\newblock In Floreano et~al. \cite{Floreano-etal99}, page 104.

\bibitem{Standish00b}
Russell~K. Standish.
\newblock Evolution in the {Multiverse}.
\newblock {\em Complexity International}, 7, 2000.

\bibitem{Standish01a}
Russell~K. Standish.
\newblock On complexity and emergence.
\newblock {\em Complexity International}, 9, 2001.

\end{thebibliography}

\end{document}